\documentclass{PoS}

\usepackage{amsfonts}
\usepackage{mathrsfs}
\usepackage{graphicx}
\usepackage{amsmath}
\usepackage{amssymb}
\usepackage{bm}
\usepackage{bbm}
\usepackage{epsfig}
\usepackage{multirow}
\usepackage{float}
\usepackage{array}
\usepackage{tabularx}
\usepackage{xcolor}

\usepackage{bm,bbm,amssymb,graphicx}
\newcommand{\tensor}[1]{\stackrel{\leftrightarrow}{#1}}

\newcommand{\nn}{\nonumber}

\newcommand{\bqa}{\begin{eqnarray}}
\newcommand{\eqa}{\end{eqnarray}}

\title{Some recent progress in understanding exclusive double charmonium production at
$\bm{B}$ factories}

\ShortTitle{Some recent progress in understanding exclusive double
charmonium production at $B$ factories}

\author{Hai-Rong Dong\\
       Institute of High Energy Physics,
       Chinese Academy of
       Sciences, Beijing 100049, China\\
        E-mail: \email{donghr@ihep.ac.cn}}

\author{Feng Feng\\
        Center for High Energy Physics, Peking University,
        Beijing 100871, China\\
        E-mail: \email{fengf@ihep.ac.cn}}

\author{\speaker{Yu Jia}\\
       Institute of High Energy Physics,
       Chinese Academy of
       Sciences, Beijing 100049, China\\
       Theoretical Physics Center for Science Facilities,
CAS, Beijing 100049, China\\
        E-mail: \email{jiay@ihep.ac.cn}}

\abstract{We review some recent progress in understanding various
exclusive double charmonium production processes at $B$ factories,
within the nonrelativistic QCD factorization framework. First we
investigate the impact of the joint perturbative and relativistic
correction on the process that has attracted a great amount of
attention in the past decade, $e^+e^-\to J/\psi+\eta_c$. We then
briefly discuss the phenomenological implication of the
next-to-leading order perturbative correction to the processes
$e^+e^-\to J/\psi+\chi_{c0,1,2} (\eta_{c2})$. We further emphasize a
novel theoretical challenge, which is recently discovered by
applying the NRQCD factorization approach to the helicity-suppressed
hard exclusive reactions involving heavy quarkonium.}

\FullConference{Xth Quark Confinement and the Hadron Spectrum\\
                 8-12 October 2012\\
                 TUM Campus Garching, Munich, Germany}

\begin{document}

\section{Introduction}
\label{Introduction}

The discovery of a number of double-charmonium production processes
at $B$ factories about a decade
ago~\cite{Abe:2002rb,Abe:2004ww,Aubert:2005tj} has triggered the
long-lasting interest. Experimentally, this production mechanism
offers a unique environment to search for the new $C$-even
charmonium states, particularly those $X$, $Y$, $Z$ states, by
fitting the recoil mass spectrum against the $J/\psi$ ($\psi'$). The
famous examples are the establishing of the $X(3940)$ and $X(4160)$
states along this avenue.

On the theoretical side, the double-charmonium production provides a
powerful arsenal to strengthen our understanding toward perturbative
QCD, especially toward the application of the light-cone
approach~\cite{Lepage:1980fj,Chernyak:1983ej} and the
nonrelativistic QCD (NRQCD) factorization
approach~\cite{Bodwin:1994jh} to hard exclusive reactions involving
heavy quarkonium.

Thanks to the very clean $J/\psi(\psi^\prime)\to l^+l^-$ signals,
experimentally it is most favorable to reconstruct those
double-charmonium events that involve a $J/\psi$ ($\psi^\prime$)
meson. Thus far, the most intensively-studied double charmonium
production process is $e^+e^-\to J/\psi+\eta_c$. The lowest-order
(LO) NRQCD predictions to this
process~\cite{Braaten:2002fi,Liu:2002wq,Hagiwara:2003cw} was about
one order of magnitude smaller than the original \textsf{Belle}
measurement~\cite{Abe:2002rb}. This disquieting discrepancy has
spurred a great amount of theoretical investigations in both NRQCD
and light-cone approaches~\cite{Brambilla:2010cs}. One crucial
ingredient in alleviating the discrepancy between the NRQCD
prediction and the data is the {\it positive} and {\it substantial}
next-to-leading order (NLO) perturbative
corrections~\cite{Zhang:2005cha,Gong:2007db}~\footnote{The
tree-level relativistic corrections to this process have also been
investigated~\cite{Braaten:2002fi,He:2007te,Bodwin:2007ga}.
Including these corrections appears to be helpful to further reduce
the gap between the NRQCD prediction and the data.}. By contrast,
owing to some long-standing theoretical obstacles for the
helicity-flipped process, {\it e.g.} the ``endpoint singularity
problem", the $O(\alpha_s)$ correction to this process has never
been worked out in the light-cone approach. Therefore, despite some
shortcomings, the NRQCD factorization approach remains to be the
{\it only} viable formalism to tackle double-charmonium production
which is both based on the first principles of QCD and also amenable
to the systematical improvement.

In this talk, we review some recent investigations on the higher
order corrections to the double-charmonium production processes
$e^+e^-\to J/\psi+\chi_{c0,1,2}(\eta_{c2})$ at $B$
factories~\cite{Wang:2011qg,Dong:2011fb,Dong:2012xx,Li:2013qp,Dong:2013qw}.
In particular, we address the impact of the $O(\alpha_s)$ [or
$O(\alpha_s v^2)$, where $v$ denotes the characteristic velocity of
the charm quark inside charmonium] corrections on these processes in
NRQCD factorization approach. The theoretical predictions will be
confronted with the measurements whenever possible. We also remark
on a novel theoretical phenomenon, {\it i.e.}, the occurrence of the
double logarithms in the NRQCD short-distance coefficients at NLO in
$\alpha_s$.

The rest of the paper is structured as follows.
In Sec.~\ref{hsr:red:hel:ampl}, we recall the helicity selection
rule relevant for the exclusive double charmonium production
processes considered in this work.
In Sec.~\ref{technique}, we sketch some key techniques underlying
the $O(\alpha_s)$ calculation.
In Sec.~\ref{alphas:v2:jpsi:etac}, we consider the $O(\alpha_s v^2)$
correction to the process $e^+ e^- \to J/\psi+\eta_c$, presenting
the asymptotic expressions of the $O(\alpha_s)$ NRQCD short-distance
coefficients through relative order-$v^2$. The phenomenological
impact of this new correction is also explored.
In Sec.~\ref{alphas:jpsi:chicj:etac2}, we briefly discuss the
effects of the $O(\alpha_s)$ corrections to the processes $e^+ e^-
\to J/\psi + \chi_{c0,1,2}(\eta_{c2})$, and propose that these types
of processes may be used to unravel the quantum number of the famous
$X(3872)$ meson.
Finally in Sec.~\ref{summary}, we present our personal perspective
on those most important problems remaining in this field, which
urgently await the exploration.


\section{Helicity selection rule for double-charmonium production}
\label{hsr:red:hel:ampl}

For the exclusive double-charmonium production processes, it is most
informative to look into the polarized cross section, where the
power-law scaling in each helicity configuration is governed by the
famous {\it helicity selection rule} (HSR)~\cite{Brodsky:1981kj}. In
the hard-scattering limit $\sqrt{s}\gg m_c\gg \Lambda_{\rm QCD}$
($\sqrt{s}$ stands for the center-of-mass energy of the $e^+e^-$
collider, $m_c$ for the charm quark mass, and $\Lambda_{\rm QCD}$
for the intrinsic QCD scale), the HSR implies that the asymptotic
behavior for the rate of producing $J/\psi$ together with a $C$-even
charmonium $H$ (with the leading $c\bar{c}$ Fock component
possessing the quantum number ${}^{2S+1} L_J^{(1)}$) in a definite
helicity configuration is~\cite{Braaten:2002fi}
\bqa
{\sigma[e^+e^-\rightarrow J/\psi(\lambda_1) + H(\lambda_2)]\over
\sigma[e^+e^-\to \mu^+\mu^-]} & \sim & v^{6+2L} \left({m_c^2\over
s}\right)^{2+|\lambda_1+\lambda_2|},
\label{helicity:selection:rule}
\eqa
where $\lambda_1$, $\lambda_2$ represent the helicities carried by
the $J/\psi$ and $H$, respectively.
Eq.~(\ref{helicity:selection:rule}) implies that the helicity state
which exhibits the slowest asymptotic decrease, thus constitutes the
``leading-twist" contribution, {\it i.e.}, $\sigma\sim 1/s^3$, is
$(\lambda_1,\lambda_2)=(0,0)$. For the processes exemplified by
$e^+e^-\to J/\psi+\eta_c(\chi_{c1},\eta_{c2})$, parity invariance
forbids the occurrence of the $(0,0)$ configuration, therefore they
are entirely of the ``higher twist" ({\it helicity-flipped}) nature.

\section{Techniques in calculating the $\bm{O}\bm{(}\bm{\alpha}_{\bm s}\bm{)}$ corrections}
\label{technique}

\begin{figure}[tbH]
\begin{center}
\includegraphics[scale=0.4]{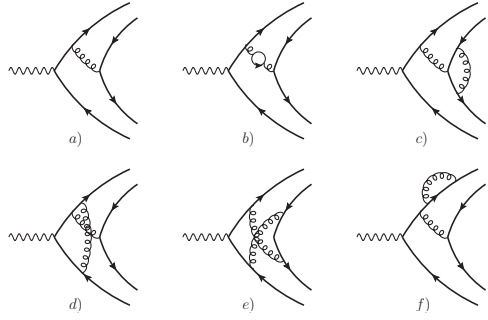}
\caption{One sample LO diagram and five sample NLO diagrams that
contribute to $\gamma^*\to J/\psi+H$, where $H$ is a $C$-even
charmonium state such as $\eta_c$, $\chi_{c0,1,2}$ or $\eta_{c2}$.
\label{feynman:diagrams}}
\end{center}
\end{figure}

The NRQCD short-distance coefficient can be most easily obtained by
computing the on-shell quark amplitude $\gamma^*\to c\bar{c} (P_1,
^3S_1^{(1)})+ c\bar{c} (P_2, ^{2S+1} L_J^{(1)})$, with the aid of
the covariant spin projectors~\cite{Bodwin:2002hg,Braaten:2002fi}.
At LO in $\alpha_s$, there are 4 diagrams for this parton process;
while at NLO in $\alpha_s$, there are 20 two-point, 20 three-point,
18 four-point, and 6 five-point one-loop diagrams. Some typical
diagrams are shown in Fig.~\ref{feynman:diagrams}.

In spite of resorting to perturbative matching method, a much more
economic way is to directly extract the short-distance coefficients
following the philosophy of {\it threshold
expansion}~\cite{Beneke:1997zp}. That is, when projecting each
$c\bar{c}$ pair onto the desired orbital-angular-momentum state, one
first expands the amplitude in powers of the relative quark momentum
$q$ prior to performing the loop integration.

Consequently, a technical complication arises, that one inevitably
encounters some unusual one-loop integrals that contain the
propagators of (up to) cubic power, due to taking the derivative
over $q$. The \textsc{Mathematica} packages
\textsc{FIRE}~\cite{Smirnov:2008iw} and the code
\textsc{Apart}~\cite{Feng:2012iq} are utilized to reduce these
unconventional higher-point one-loop tensor integrals into a minimal
set of masters integrals. Thanks to the integration-by-part
algorithm and partial fraction technique built into these codes, it
turns out that all the encountered master integrals are nothing but
the ordinary 2-point and 3-point one-loop scalar integrals, whose
analytic expressions can be found in Appendix of
Ref.~\cite{Gong:2007db}.

When adding the contributions of all the diagrams, and after
renormalizing the charm quark mass and the QCD coupling constant, we
end up with both UV and IR finite NLO expressions for the
$O(\alpha_s )$ amplitude associated with $\gamma^*\to
J/\psi+\chi_{c0,1,2}(\eta_{c2})$. In comparison, the amplitude for
$\gamma^*\to J/\psi+\eta_c$ contains an uncanceled IR pole at
$O(\alpha_s v^2)$. Fortunately, it can be factored into the relative
order-$v^2$ NRQCD matrix element via the pull-up mechanism. In
either case, one ends up with the IR-finite $O(\alpha_s)$ NRQCD
short-distance coefficients for all the helicity amplitudes
affiliated with $e^+e^-\to J/\psi+H$. Our calculation explicitly
confirms the assertion made in Ref.~\cite{Bodwin:2008nf}, that the
NRQCD factorization holds beyond tree level for the exclusive
production of a $S$-wave quarkonium plus any higher
orbital-angular-momentum quarkonium in $e^+e^-$ annihilation.

\section{${\bm{O}}\bm{(}\bm{\alpha}_{\bm{s}}
\bm{v}^{\bm{2}}\bm{)}$ correction to $\bm{e}^{\bm+}
\bm{e}^{\bm-}\bm{\to} \bm{J}\bm{/}\bm{\psi}\bm{+}\bm{\eta}_{\bm c}$}
\label{alphas:v2:jpsi:etac}

The production rate for $e^+e^-\to J/\psi + \eta_c$ can be expressed
as
\bqa
& & \sigma[e^+e^-\to J/\psi + \eta_c] = {4\pi \alpha^2 \over 3}
\left({|{\bf P}| \over \sqrt{s}}\right)^3 \left|G(s)\right|^2,
\label{integrated:cross:section:Jpsi:etac}
\eqa
where $|{\bf P}|$ signifies the magnitude of the momentum carried by
the $J/\psi$ ($\eta_c$) in the center-of-mass frame. $G(s)$ is the
$J/\psi+\eta_c$ time-like electromagnetic form factor, defined
through
$\langle J/\psi(P_1,\lambda)+\eta_c(P_2)\vert J_{\rm EM}^\mu \vert 0
\rangle = i\,G (s)\,\epsilon^{\mu\nu\rho\sigma} P_{1\nu} P_{2 \rho}
\varepsilon^*_\sigma (\lambda)$, where $J^\mu_{\rm EM}$ is the
electromagnetic current, $P_1$ ($\varepsilon^\sigma (\lambda)$)
denote the momentum (polarization vector) of the $J/\psi$, and $P_2$
the momentum of the $\eta_c$, respectively.

\subsection{NRQCD factorization formula}

NRQCD factorization allows one to factorize the $J/\psi+\eta_c$
electromagnetic form factor as
\bqa
\label{EMFF:NRQCD:factorization}
G(s) &=&  \sqrt{4 M_{J/\psi} M_{\eta_c}}\langle J/\psi| \psi^\dagger
\bm{\sigma} \cdot \bm{\epsilon} \chi |0\rangle \langle \eta_c|
\psi^\dagger \chi | 0 \rangle  \left[c_0 + c_{2,1} \langle v^2
\rangle_{J/\psi} + c_{2,2} \langle v^2 \rangle_{\eta_c} + \cdots
\right],
\eqa
where $c_0$ and $c_{2,i}$ are the dimensionless short-distance
coefficients that depend on $m_c^2/s$. For simplicity, we have also
introduced the following dimensionless ratios of NRQCD matrix
elements to signify the $O(v^2)$ corrections:
$\langle v^2 \rangle_{J/\psi} = \langle J/\psi(\lambda)|
\psi^\dagger (-\tfrac{i}{2}\tensor{\mathbf{D}})^{2}
\bm{\sigma}\cdot\bm{\epsilon}(\lambda)\chi|0\rangle/(m_c^2\, \langle
J/\psi(\lambda)| \psi^\dagger
\bm{\sigma}\cdot\bm{\epsilon}(\lambda)\chi|0\rangle)$,
$ \langle v^2 \rangle_{\eta_c} = \langle \eta_c| \psi^\dagger
(-\tfrac{i}{2}\tensor{\mathbf{D}})^{2} \chi|0\rangle/(m_c^2\,
\langle \eta_c| \psi^\dagger \chi|0\rangle)$, where $\psi^\dagger
\tensor{\mathbf{D}}\chi\equiv \psi^\dagger {\mathbf{D}} \chi-
({\mathbf{D}} \psi)^\dagger \chi$.

Inserting Eq.~(\ref{EMFF:NRQCD:factorization}) into
(\ref{integrated:cross:section:Jpsi:etac}), one can decompose the
cross section into the $O(v^0)$ and $O(v^2)$ pieces,
$ \sigma [e^+e^-\to J/\psi+\eta_c] = \sigma_0+\sigma_2+ O(\sigma
v^4)$,
where
\begin{subequations}
\bqa
\sigma_0 & = & {8\pi\alpha^2 m_c^2 (1-4 r)^{3/2}\over 3} \langle
{\mathcal O}_1 \rangle_{J/\psi} \langle{\mathcal
O}_1\rangle_{\eta_c}|c_{0}|^2,
\\
\sigma_2 & = & {4\pi \alpha^2 m_c^2 (1-4 r)^{3/2}\over 3} \langle
{\mathcal O}_1 \rangle_{J/\psi} \langle{\mathcal
O}_1\rangle_{\eta_c}
\label{sigma_2:definition}
\\
&& \Bigg\{ \bigg( {1-10r \over 1-4r} |c_{0}|^2
 + 4\,{\rm Re}[c_{0} c_{2,1}^*]
\bigg) \langle v^2\rangle_{J/\psi}
  + \bigg( {1-10 r \over 1-4r} |c_{0}|^2 + 4\,{\rm Re}[c_{0}
c_{2,2}^*] \bigg ) \langle v^2\rangle_{\eta_c} \Bigg\}.\nn
\eqa
\label{sigma0:sigma2:definitions}
\end{subequations}
We have employed the Gremm-Kapustin relation~\cite{Gremm:1997dq} to
eliminate the explicit occurrences of $M_{J/\psi}$ and $M_{\eta_c}$
in (\ref{sigma0:sigma2:definitions}). To condense the notation, we
have introduced the following symbols: $r=4 m_c^2/s$, $\langle
\mathcal{O}_1\rangle_{J/\psi}=\big| \langle
J/\psi({\bm{\epsilon}})|\psi^\dag \bm{\sigma} \cdot \bm{\epsilon}
\chi |0\rangle \big|^2$, and $\langle {\mathcal O}_1
\rangle_{\eta_c} = \big| \langle {\eta_c}| \psi^\dag \chi |0\rangle
\big|^2$.

\subsection{Various NRQCD short-distance coefficients}

We organize the coefficients $c_i$ ($i=0,2$) in power series of the
strong coupling constant, {\it i.e.}, $c_i=c_i^{(0)}+{\alpha_s\over
\pi}c_i^{(1)}+\cdots$. Accordingly, one may decompose the cross
section $\sigma_i$ into $\sigma_i^{(0)}+\sigma_i^{(1)}$ ($i=0,2$) as
well. Thus far, the only missing piece is $\sigma_2^{(1)}$. The
tree-level short-distance coefficients through $O(v^2)$ have been
available long ago~\cite{Braaten:2002fi}:
\bqa
c_{0}^{(0)}&=& {32 \pi C_F e_c \alpha_s \over N_c\,m_c s^2},
\qquad
c_{2,1}^{(0)}=  \frac{3-10r}{6} c_{0}^{(0)},
\qquad c_{2,2}^{(0)}= \frac{2-5r}{3} c_{0}^{(0)},
\label{c_i:tree-level:values}
\eqa
where $e_c={2\over 3}$ is the electric charge of the charm quark,
and $C_F={N_c^2-1\over 2N_c}={4\over 3}$.

The $O(\alpha_s)$ NRQCD short-distance coefficients $c_0^{(1)}$ and
$c_{2,i}^{(1)}$ are generally complex-valued, and
cumbersomely-looking. Nevertheless, it is much more illuminating to
look at their asymptotic expressions in the limit $\sqrt{s}\gg m_c$:
\begin{subequations}
\bqa
& & c_{0}^{(1)}\left( r,{\mu_r^2 \over s} \right)_{\rm asym} =
c_{0}^{(0)}\times \Bigg\{ \beta_0 \bigg(- \frac{1}{4}
\ln\frac{s}{4\mu_r^2} + \frac{5}{12}\bigg) + \bigg( \frac{13}{24}
\ln^2 r + \frac{5}{4} \ln2 \ln r - \frac{41}{24} \ln r
\nn\\
&&- \frac{53}{24}\ln^2 2+\frac{65}{8}\ln2
-\frac{1}{36}\pi^2-\frac{19}{4} \bigg)
 + i\pi \bigg( \frac{1}{4}\beta_0 + \frac{13}{12}\ln r  +
\frac{5}{4} \ln2- \frac{41}{24} \bigg)  \Bigg\},
\label{c_0:NLO:asym:expressions}
\\
& & c_{2,1}^{(1)}\left( r,{\mu_r^2 \over s},{\mu_f^2 \over m_c^2}
\right)_{\rm asym} = {1\over 2} c_{0}^{(0)}\times \Bigg\{
\frac{16}{9}\ln\frac{\mu_f^2}{m_c^2} + \beta_0 \bigg( -\frac{1}{4}
\ln\frac{s}{4\mu_r^2} + \frac{11}{12}\bigg) + \bigg( \frac{3}{8}
\ln^2 r + \frac{19}{12} \ln2 \ln r
\nn\\
&& + \frac{31}{24} \ln r- \frac{1}{24}\ln^2 2
+\frac{893}{216}\ln2-\frac{5}{36}\pi^2-\frac{497}{72} \bigg)
 + i\pi \bigg( \frac{1}{4}\beta_0 + \frac{3}{4}\ln r +
\frac{19}{12} \ln2 + \frac{9}{8} \bigg)  \Bigg\},
\\
&& c_{2,2}^{(1)}\left( r,{\mu_r^2 \over s},{\mu_f^2 \over m_c^2}
\right)_{\rm asym} =  {2\over 3} c_{0}^{(0)}\times  \Bigg\{
\frac{4}{3}\ln\frac{\mu_f^2}{m_c^2} +  \beta_0 \bigg( -\frac{1}{4}
\ln\frac{s}{4\mu_r^2} + \frac{2}{3}\bigg) + \bigg( \frac{1}{12}
\ln^2 r + \frac{11}{12} \ln2 \ln r
\nn\\
&&  - \frac{1}{24} \ln r - \frac{11}{8}\ln^2
2+\frac{241}{144}\ln2-\frac{1}{8}\pi^2 - \frac{99}{16} \bigg)
 + i\pi \bigg( \frac{1}{4} \beta_0 + \frac{1}{6}\ln r +
\frac{11}{12} \ln2 - \frac{1}{24} \bigg)  \Bigg\},
\eqa
\label{c_i:NLO:asym:expressions}
\end{subequations}
where $\beta_0={11\over 3} C_A-{2\over 3}n_f$ is the one-loop
coefficient of the QCD $\beta$ function, and $n_f=4$ denotes the
number of active quark flavors. $\mu_r$ denotes the renormalization
scale, and $\mu_f$ signifies the NRQCD factorization scale in the
$\overline{\rm MS}$ scheme, which naturally ranges from $m_c v$ to
$m_c$.

As first pointed out in Ref.~\cite{Jia:2010fw}, a peculiar
double-logarithmic correction $\propto \ln^2 r$ arises in
$c_0^{(1)}$ for this helicity-flipped process, and our
(\ref{c_0:NLO:asym:expressions}) exactly agrees with the
corresponding expression there~\footnote{Our result slightly differs
from Ref.~\cite{Gong:2007db} on the imaginary part of $c_0^{(1)}$,
though it does not affect the phenomenology.}.
Eqs.~(\ref{c_i:NLO:asym:expressions}) imply that the double
logarithms survive at $O(\alpha_s v^2)$ as well. With $\sqrt{s}\gg
m_c$, one presumably needs to sum these types of logarithms to all
orders in $\alpha_s$ to warrant a reliable prediction. Such a
resummation may even be mandatory at the $B$-factory
energy~\cite{Jia:2010fw}. At present, how to fulfill this goal
remains to be a thorny challenge.

\subsection{Phenomenological impact of the $\bm{O}\bm{(}\bm{\alpha}_{\bm{s}}
\bm{v}^{\bm{2}}\bm{)}$ correction}
\label{phenomenology}

Apart from the ambiguity in the values of $m_c$ and NRQCD matrix
elements, the freedom of choosing the scale entering the strong
coupling constant leads to a large uncertainty for our prediction to
$\sigma[e^+e^-\to J/\psi+\eta_c]$. This is a serious drawback of the
NRQCD approach~\cite{Jia:2008ep}. For simplicity we assign all the
occurring $\alpha_s$ with a common scale, $\mu_r$, and choose
$\mu_r=\sqrt{s}/2$ and $\mu_r=2 m_c$, respectively, hoping that the
less biased results interpolate somewhere in between.

\begin{table}[tbH]
\caption{The individual contribution to the predicted
$\sigma[e^+e^-\to J/\psi+\eta_{c}]$ at $\sqrt{s}=10.58$ GeV, as
specified by the powers of $\alpha_s$ and $v^2$. We take
$\alpha(\sqrt{s}) = 1/130.9$, $\Lambda^{(4)}_{\overline{\rm MS}}=
0.338$ GeV, $m_c=1.4$ GeV, and $\mu_f=m_c$. The LO NRQCD matrix
elements are $ \langle {\mathcal{O}_1}\rangle_{J/\psi}\approx\langle
{\mathcal{O}_1}\rangle_{\eta_c}=0.387\;{\rm GeV}^3 $. The
Gremm-Kapustin relation is used to obtain $\langle
v^2\rangle_{J/\psi} = 0.223$ and $\langle v^2\rangle_{\eta_c} =
0.133$. The cross sections are in units of fb.
\label{tbl:cross:component}}
\begin{center}
\begin{tabular}{|c|c|c|c|c|}
\hline
$\alpha_s(\mu_r)$  &  $\sigma_0^{(0)}$  & $\sigma_0^{(1)}$ & $\sigma_2^{(0)}$& $\sigma_2^{(1)}$ \\
\hline
   $\alpha_s(\frac{\sqrt{s}}{2}) = 0.211$  &  $4.40$  & $5.22$  & $1.72$ & $0.73$ \\
\hline
   $\alpha_s(2 m_c) = 0.267$  &  $7.00$  & $7.34$  & $2.73$ & $0.24$ \\
\hline
\end{tabular}
\end{center}
\end{table}

Table~\ref{tbl:cross:component} lists the predicted
$\sigma[e^+e^-\to J/\psi+\eta_{c}]$ with two sets of $\mu_r$,
organized in double expansions of $\alpha_s$ and $v$. We reproduce
the well-known results, i.e., the positive and substantial
$O(\alpha_s)$ correction~\cite{Zhang:2005cha,Gong:2007db}, and the
positive but less pronounced $O(v^2)$
correction~\cite{Braaten:2002fi,He:2007te,Bodwin:2007ga}. Taking
$r=0.07$, which is relevant for the $B$ factory energy, ${\rm
Re}[c_{2,i}^{(1)}/c_{2,i}^{(0)}]$ ($i=1, 2$) turn out to be both
large and negative. One might naively expect that including the new
$ O(\alpha_s v^2)$ correction would largely dilute the existing
$O(v^2)$ term. However, the new correction to the cross section,
$\sigma_2^{(1)}$, is actually positive and modest. This may be
attributed to the accidental cancelation between the two terms in
the prefactor of $\langle v^2 \rangle_{H}$ in
(\ref{sigma_2:definition}), which represent two difference sources
of relativistic correction.

\begin{figure}[tbH]
\begin{center}
\includegraphics[width=0.4\textwidth]{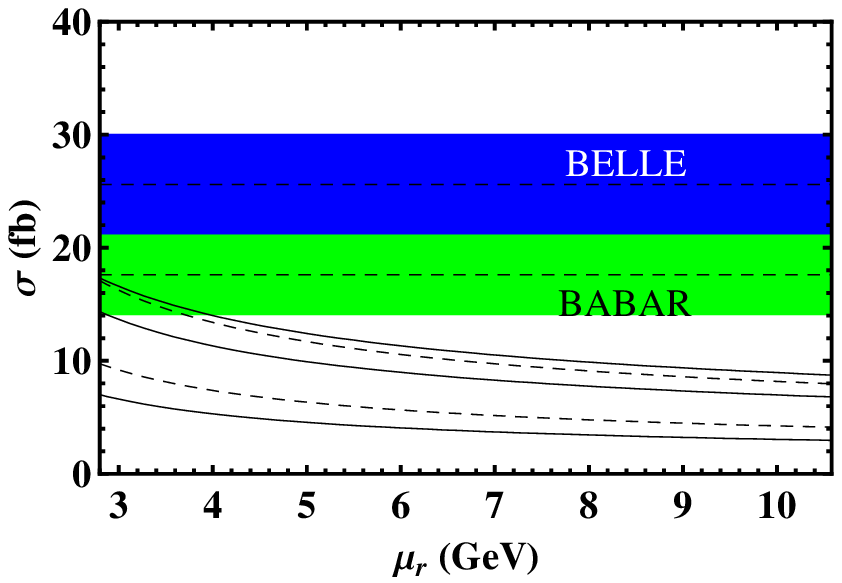}
\includegraphics[width=0.42\textwidth]{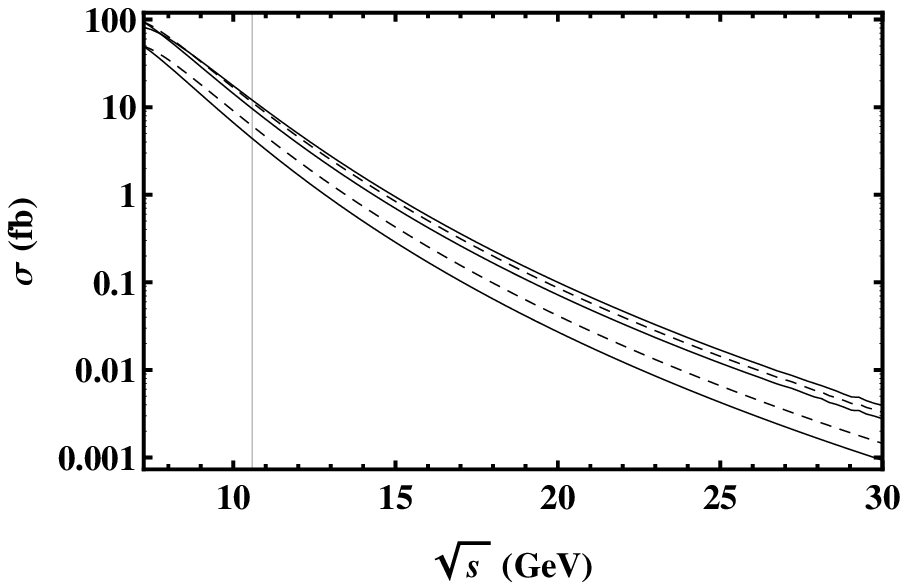}
\caption{The $\mu_r$- and $\sqrt{s}$-dependence of the cross section
for $e^+e^-\to J/\psi+\eta_{c}$. The 5 curves from bottom to top are
$\sigma_0^{(0)}$ (solid line), $\sigma_0^{(0)}+\sigma_2^{(0)}$
(dashed line), $\sigma_0^{(0)}+\sigma_0^{(1)}$ (solid line),
$\sigma_0^{(0)}+\sigma_2^{(0)}+\sigma_0^{(1)}$ (dashed line), and
$\sigma_0^{(0)}+\sigma_2^{(0)}+\sigma_0^{(1)}+\sigma_2^{(1)}$ (solid
line), respectively. In the left panel, we fix $\sqrt{s}=10.58$ GeV,
and the blue and green bands represent the measured cross sections
by the \textsf{Belle} and \textsf{BaBar} experiments, with
respective systematic and statistical errors added in quadrature.
\label{crossX:figure} }
\end{center}
\end{figure}

In Fig.~\ref{crossX:figure}, we plot the $\sigma[e^+e^-\to
J/\psi+\eta_{c}]$ as a function of $\mu_r$ and $\sqrt{s}$. At the
$B$ factory energy, incorporating the new correction
$\sigma_2^{(1)}$ appears not to make a big difference. When $\mu_r$
is relatively small, the state-of-the-art NRQCD prediction converges
to the \textsf{BaBar} measurement within errors. Had we taken
somewhat larger values of the NRQCD matrix elements $\langle
{\mathcal{O}_1}\rangle_{H}$ as in \cite{He:2007te,Bodwin:2007ga},
the agreement with the $B$ factory measurements would improve.

As shown in the right panel of Fig.~\ref{crossX:figure},
$\sigma[e^+e^-\to J/\psi+\eta_{c}]$ drops steeply as $\sqrt{s}$
increases, reflecting the helicity-flipped nature of this process,
$\sigma\sim 1/s^4$. In contrast to its minor impact at $B$ factory,
the $O(\alpha_s v^2)$ correction turns out to be much more relevant
at higher $\sqrt{s}$.

\section{$\bm{O}\bm{(} \bm{\alpha}_{\bm{s}}\bm{)}$ correction to
$\bm{e}^{\bm +} \bm{e}^{\bm -}\bm{\to}
\bm{J}\bm{/}\bm{\psi}\bm{+}\bm{\chi}_{\bm{c}\bm{0}\bm{,}\bm{1}\bm{,}
\bm{2}} \bm{(}\bm{\eta}_{\bm{c}\bm{2}}\bm{)}$}
\label{alphas:jpsi:chicj:etac2}

The NLO perturbative corrections to the double-charmonium production
processes $e^+e^-\to J/\psi+\chi_{c0,1,2}(\eta_{c2})$ have also been
investigated recently~\cite{Wang:2011qg,Dong:2011fb,Dong:2013qw}.
The calculational technique resembles that for the $O(\alpha_s v^2)$
correction to $e^+e^-\to J/\psi+\eta_{c}$, and has been reviewed in
Sec.~\ref{technique}. The processes $e^+e^-\to J/\psi+\chi_{c0,1,2}$
have been independently studied by two groups, and they agree with
each other~\cite{Wang:2011qg,Dong:2011fb}.

The $O(\alpha_s)$ correction to $e^+e^-\to J/\psi+\chi_{c0}$ appears
to be substantial. With some reasonable input parameters, and the
renormalization scale $\mu_r$ set as ${\sqrt{s}\over 2}$, including
the $O(\alpha_s)$ correction enhances the LO cross section from 4.8
fb to 8.6 fb. This value is compatible with the \textsf{Belle} and
\textsf{BaBar} measurements within errors. However, the NLO
perturbative prediction to the $\psi'+\chi_{c0}$ production rate is
still significantly below the central value of the \textsf{Belle}
measurement~\cite{Wang:2011qg,Dong:2011fb}.

The situation for $e^+e^-\to J/\psi+\chi_{c1,2}$ is less clear. The
impact of the NLO perturbative corrections to these processes seems
to be modest, even with the sign uncertain, depending on the
different choices of $\mu_r$. The predicted cross sections for both
processes are around 1 fb, which are almost one order of magnitude
smaller than that for $e^+e^-\to J/\psi+\chi_{c0}$. Encouragingly,
the recent \textsf{Belle} experiment~\cite{Pakhlov:2009nj} did
observe a considerable number of $e^+e^-\to J/\psi+\chi_{c1,2}$
events, from which one may roughly estimate the corresponding
production cross sections. They appear to be qualitatively
consistent with our expectations.

Recently there has arisen some controversy about the canonical
charmonium option of the $X(3872)$ meson, whether it being
$\eta_{c2}$ or $\chi^\prime_{c1}$~\cite{delAmoSanchez:2010jr}.
Motivated by this concern, we have also performed a comparative
study for $e^+e^-\to J/\psi+\eta_{c2}$ and $e^+e^-\to
J/\psi+\chi^\prime_{c1}$ at $B$-factory energy, hoping that it may
provide some guidance to unravel the quantum number of the $X(3872)$
meson in the future experiments~\cite{Dong:2013qw}. The NLO
perturbative correction to the former process is of medium size.
With $\mu_r$ taken as $\sqrt{s}/2$, implementing the $O(\alpha_s)$
correction enhances the LO cross section from 0.22 fb to 0.29 fb.
The production rate of the latter process is about 6-7 times greater
than the former, thereby it seems realistic to observe the
$J/\psi+\chi^\prime_{c1}$ signals based on the current 1 ${\rm
ab}^{-1}$ \textsc{Belle} data sample, if the $\chi^\prime_{c1}$ is
indeed the narrow $X(3872)$ meson.

Refs.~\cite{Dong:2011fb,Dong:2013qw} also conduct a comprehensive
study on the polarized cross sections for $e^+e^-\to
J/\psi+\chi_{c0,1,2}(\eta_{c2})$. It is found that the bulk of the
total cross section comes from the $(0,\pm 1)$ helicity channels for
$e^+e^-\to J/\psi+\chi_{c1}$, from the $(0,0)$ and $(\pm 1,0)$
helicity states for $e^+e^-\to J/\psi+\chi_{c2}$, and from the $(\pm
1,0)$ states for $e^+e^-\to J/\psi+\eta_{c2}$. The hierarchy among
the various helicity channels appears to often conflict with what is
expected from the HSR. It will be interesting for the future
experiments to concretely test these polarization patterns.

By working out the asymptotic expressions of the various helicity
amplitudes for the processes $\gamma^*\to
J/\psi+\chi_{c0,1,2}(\eta_{c2})$, we firmly confirm the pattern
speculated in Ref.~\cite{Jia:2010fw}: The hard exclusive reaction
involving double charmonium at leading twist can only host the
single collinear logarithm $\ln{s/m_c^2}$ at NLO in $\alpha_s$,
while the double logarithms of form $\ln^2{s/m_c^2}$ are always
associated with those helicity-suppressed channels.

\section{Outlook}
\label{summary}

After a decade of intensive study, our understanding of exclusive
double charmonium production has gradually matured. The most notable
lesson is perhaps that, the NRQCD factorization approach has proved
to be a successful and indispensable tool in dealing with hard
exclusive reactions involving heavy quarkonium. However, in our
opinion, this research area is still far from being closed, and
there remain some important questions to be answered. In the
following, we enumerate two topics which may urgently beg for the
exploration.

We have reviewed some recent advances in the NLO perturbative
correction to double charmonium production processes. Aside from
$e^+e^-\to J/\psi+\eta_c$, the relativistic correction has hardly
been investigated for any double-charmonium production process
involving the $P$, $D$-wave charmonium. There is no ground to
believe their effects are less important than the perturbative
corrections. A practical difficulty to assess the relativistic
correction for these processes is that, in general more NRQCD matrix
elements than for the $S$-wave charmonium will come into play, about
whose values we have almost no any clue. Hopefully, the lattice
NRQCD simulation will eventually provide some useful information for
those long-distance matrix elements.

A great theoretical challenge is to tame the double logarithms of
form $\ln^2(s/m_c^2)$ in the $O(\alpha_s)$ NRQCD short-distance
coefficients which are always affiliated with the
helicity-suppressed exclusive quarkonium production channels. The
occurrence of these process-dependent, (positive) double logarithms
severely jeopardizes the reliability of the fix-order perturbation
theory prediction. We note that some important progress has been
made recently in tracing the origin of these double logarithms at
one-loop order (differentiating the harmless Sudakov double
logarithm from the problematic endpoint double
logarithm)~\cite{Bodwin:2013ys}. Nevertheless, there is still a long
way to go to finally develop a systematic control over these
endpoint double logarithms, {\it e.g.}, to resum them to all orders
in $\alpha_s$. In our perspective, the occurrence of these double
logarithms is likely intertwined with the long-standing failure in
applying the light-cone approach to the hard exclusive reactions
beyond tree level. Looking on the bright side, with explicit
expressions of the $O(\alpha_s)$ NRQCD short-distance coefficients
for many channels at our disposal, one may view the exclusive double
quarkonium production as a fertile theoretical laboratory, from
which some fresh insight may be gained by reexamining those old
problems of the light-cone approach.

\acknowledgments

This research was supported in part by the National Natural Science
Foundation of China under Grant Nos.~10935012, 11125525, DFG and
NSFC (CRC 110), and by the Ministry of Science and Technology of
China under Contract No. 2009CB825200.


\begin{thebibliography}{99}

\bibitem{Abe:2002rb}
  K.~Abe {\it et al.}  [Belle Collaboration],
  Phys.\ Rev.\ Lett.\  {\bf 89}, 142001 (2002),
  [arXiv:hep-ex/0205104].

\bibitem{Abe:2004ww}
  K.~Abe {\it et al.}  [Belle Collaboration],
  Phys.\ Rev.\  D {\bf 70}, 071102 (2004)
  [arXiv:hep-ex/0407009].

\bibitem{Aubert:2005tj}
  B.~Aubert {\it et al.}  [BABAR Collaboration],
  Phys.\ Rev.\  D {\bf 72}, 031101 (2005)
  [arXiv:hep-ex/0506062].


\bibitem{Lepage:1980fj}
  G.~P.~Lepage and S.~J.~Brodsky,
  Phys.\ Rev.\  D {\bf 22}, 2157 (1980).

\bibitem{Chernyak:1983ej}
  V.~L.~Chernyak and A.~R.~Zhitnitsky,
  Phys.\ Rept.\  {\bf 112}, 173 (1984).

\bibitem{Bodwin:1994jh}
  G.~T.~Bodwin, E.~Braaten and G.~P.~Lepage,
  Phys.\ Rev.\  D {\bf 51}, 1125 (1995)
  [Erratum-ibid.\  D {\bf 55}, 5853 (1997)]
  [arXiv:hep-ph/9407339].

\bibitem{Braaten:2002fi}
  E.~Braaten and J.~Lee,
  Phys.\ Rev.\ D {\bf 67}, 054007 (2003)
  [arXiv:hep-ph/0211085].

\bibitem{Liu:2002wq}
  K.~Y.~Liu, Z.~G.~He and K.~T.~Chao,
  Phys.\ Lett.\  B {\bf 557}, 45 (2003)
  [arXiv:hep-ph/0211181].

\bibitem{Hagiwara:2003cw}
  K.~Hagiwara, E.~Kou and C.~F.~Qiao,
  Phys.\ Lett.\  B {\bf 570} (2003) 39
  [arXiv:hep-ph/0305102].

\bibitem{Brambilla:2010cs}
 For a recent review, see N.~Brambilla {\it et al.},
  Eur.\ Phys.\ J.\  C {\bf 71}, 1534 (2011), and references therein.

\bibitem{Zhang:2005cha}
  Y.~-J.~Zhang, Y.~-j.~Gao, K.~-T.~Chao,
  Phys.\ Rev.\ Lett.\  {\bf 96}, 092001 (2006).
  [hep-ph/0506076].

\bibitem{Gong:2007db}
  B.~Gong and J.~X.~Wang,
  Phys.\ Rev.\  D {\bf 77}, 054028 (2008)
  [arXiv:0712.4220 [hep-ph]].

\bibitem{He:2007te}
  Z.~G.~He, Y.~Fan and K.~T.~Chao,
  Phys.\ Rev.\  D {\bf 75}, 074011 (2007)
  [arXiv:hep-ph/0702239].

\bibitem{Bodwin:2007ga}
  G.~T.~Bodwin, J.~Lee and C.~Yu,
  Phys.\ Rev.\  D {\bf 77}, 094018 (2008)
  [arXiv:0710.0995 [hep-ph]].

\bibitem{Wang:2011qg}
  K.~Wang, Y.~-Q.~Ma, K.~-T.~Chao,
  Phys.\ Rev.\  {\bf D84}, 034022 (2011).


\bibitem{Dong:2011fb}
  H.~-R.~Dong, F.~Feng and Y.~Jia,
  JHEP {\bf 1110}, 141 (2011)  [arXiv:1107.4351v3 [hep-ph]].


\bibitem{Dong:2012xx}
  H.~-R.~Dong, F.~Feng and Y.~Jia,
  Phys.\ Rev.\ D {\bf 85}, 114018 (2012)  [arXiv:1204.4128 [hep-ph]].

\bibitem{Li:2013qp}
  X.~-H.~Li and J.~-X.~Wang,
  arXiv:1301.0376 [hep-ph].  


\bibitem{Dong:2013qw}
  H.~-R.~Dong, F.~Feng and Y.~Jia,
  arXiv:1301.1946 [hep-ph].


\bibitem{Brodsky:1981kj}
  S.~J.~Brodsky and G.~P.~Lepage,
  Phys.\ Rev.\  D {\bf 24}, 2848 (1981).

\bibitem{Bodwin:2002hg}
  G.~T.~Bodwin and A.~Petrelli,
  Phys.\ Rev.\  D {\bf 66}, 094011 (2002)
  [arXiv:hep-ph/0205210].


\bibitem{Beneke:1997zp}
  M.~Beneke and V.~A.~Smirnov,
  Nucl.\ Phys.\  B {\bf 522}, 321 (1998)
  [arXiv:hep-ph/9711391].

\bibitem{Smirnov:2008iw}
  A.~V.~Smirnov,
  JHEP {\bf 0810}, 107 (2008).
  [arXiv:0807.3243 [hep-ph]].

\bibitem{Feng:2012iq}
  F.~Feng,
  Comput.\ Phys.\ Commun.\  {\bf 183}, 2158 (2012)  [arXiv:1204.2314 [hep-ph]].

\bibitem{Bodwin:2008nf}
  G.~T.~Bodwin, X.~Garcia i Tormo and J.~Lee,
  Phys.\ Rev.\ Lett.\  {\bf 101}, 102002 (2008).

\bibitem{Gremm:1997dq}
  M.~Gremm and A.~Kapustin,
  Phys.\ Lett.\ B {\bf 407}, 323 (1997)  [hep-ph/9701353].  

\bibitem{Jia:2010fw}
  Y.~Jia, J.~-X.~Wang and D.~Yang,
  JHEP {\bf 1110}, 105 (2011)  [arXiv:1012.6007 [hep-ph]].  

\bibitem{Jia:2008ep}
  Y.~Jia and D.~Yang,
  Nucl.\ Phys.\  B {\bf 814} (2009) 217
  [arXiv:0812.1965 [hep-ph]].

\bibitem{Pakhlov:2009nj}
  P.~Pakhlov {\it et al.}  [Belle Collaboration],
  Phys.\ Rev.\ D {\bf 79}, 071101 (2009) [arXiv:0901.2775 [hep-ex]].

\bibitem{delAmoSanchez:2010jr}
  P.~del Amo Sanchez {\it et al.}  [BABAR Collaboration],
  Phys.\ Rev.\ D {\bf 82}, 011101 (2010).

\bibitem{Bodwin:2013ys}
  G.~T.~Bodwin, H.~S.~Chung and J.~Lee,
  arXiv:1301.3937 [hep-ph], this proceeding.


\end{thebibliography}
\end{document}